\documentclass{ws-procs975x65}
\newcommand{\be}{\begin{equation}}
\newcommand{\ee}{\end{equation}}

\newcommand{\GA}{\alpha}
\newcommand{\GB}{\beta}
\newcommand{\GG}{\gamma}
\newcommand{\GD}{\delta}

\newcommand{\GK}{\kappa}

\newcommand{\GR}{\rho}

\newcommand{\GT}{\tau}

\begin{document}

\title{Flat, radiation universes with quadratic corrections and asymptotic analysis}

\author{Spiros Cotsakis$\dagger$ and Antonios Tsokaros$\ddagger$}
\address{University of the Aegean, Karlovassi, Samos 83200, Greece\\
$\dagger$\email{skot@aegean.gr, $\ddagger$atsok@aegean.gr}}

\bodymatter
\begin{abstract}\end{abstract}
It was shown long ago by T. V. Ruzmaikina and A. A. Ruzmaikin in\cite{RR} that within the framework of a homogeneous
and isotropic cosmological model quadratic corrections of the gravitational field cannot provide
solutions that are both regular initially and go over to Friedmann type at later times. We find here, by applying
a dynamical systems approach \cite{CB}, the general form of the solution to this class of models in the neighborhood of the initial singularity
 under the above conditions.
\label{sec:intro} Our starting point in this brief paper is the  Lagrangian of the general quadratic gravity theory given in the form\footnote{the
conventions for the metric and the Riemann tensor are those of \cite{LL}.}
\be L(R)=R + BR^2 + CR^{ij}R_{ij} + DR^{ijkm}R_{ijkm} .
\label{eq:lagra}
\ee
Since for a general spacetime we have the following
identity, \be \GD \int (R^2 - 4R^{ij}R_{ij} + R^{ijkm}R_{ijkm})
\sqrt{-g} d\Omega=0, \label{eq:gentity} \ee  in the derivation
of the field equations through variation of the action associated with (\ref{eq:lagra}),
only terms up to $R^{ij}R_{ij}$ will matter. If we further restrict
ourselves to isotropic spacetimes we have a second  identity of the form
\be \GD \int (R^2 - 3R^{ij}R_{ij}) \sqrt{-g} d\Omega=0 \: ,
\label{eq:isontity} \ee which enables us to keep terms only up to
$R^2$ in (\ref{eq:lagra}). In the specific model treated herein, we
consider a spatially flat universe with metric given by\be ds^2=dt^2-b(t)^2(dx^2+dy^2+dz^2),
\label{eq:flatun} \ee  assumed to be radiation dominated,
i.e. $P=\GR/3$. Under these assumptions, the variation of the action functional constructed using (\ref{eq:lagra}) gives
the following higher order field equations:
\be \frac{8\pi G}{c^4}T^{ij} =
R^{ij}-\frac{1}{2}g^{ij}R+
                           \frac{\GK}{6} \left[2RR^{ij}-\frac{1}{2}R^2g^{ij}-2(g^{ik}g^{jm}-g^{ij}g^{km})\nabla_{k}\nabla_{m}R \right]
\label{eq:fe}
\ee
where $\GK=6B+2C$.
Using (\ref{eq:isontity}), (\ref{eq:flatun}) the tt-component of (\ref{eq:fe}) can be written in the form
\be
\frac{\dot{b}^2}{b^2}-\GK\left[2\: \frac{\dddot{b}\:\dot{b}}{b^2} + 2\:\frac{\ddot{b}\dot{b}^2}{b^3}-\frac{\ddot{b}^2}{b^2} - 3\:
\frac{\dot{b}^4}{b^4} \right]
=\frac{b_{1}^2}{b^4} ,
\label{eq:beq}
\ee
where $b_{1}$ is a constant defined from
$\:\: \frac{8\pi G \rho}{3c^4}=\frac{b_{1}^2}{b^4} \:\: $ $(\nabla_{i}T^{i0}=0)$.
Note that the Friedmann solution $\sqrt{2b_1 t}$ satisfies the above equation.

Assuming that Eq. (\ref{eq:beq}) has a solution with a regular minimum at
$t=t_0$, ($\dot{b}_0=\dot{b}(t=t_0)=0$ and $b_0=b(t=t_0)\neq 0$)
we can expand this solution as a Taylor series
\be
b(t)=b_{0} + \frac{\ddot{b}_{0}}{2}\: (t-t_0)^{2} + \frac{\dddot{b}_{0}}{6}\: (t-t_0)^{3} + \cdots \:\:.
\label{eq:taylor}
\ee
Direct substitution to Eq. (\ref{eq:beq}) restricts the value of the constant
$\GK$ to $\GK=(b_{1}/(b_{0}\ddot{b}_{0}))^{2} > 0$\footnote{Due to a sign mistake,
 Ruzmaikina-Ruzmaikin in \cite{RR} conclude the opposite.}.
For this value of $\GK$, Ruzmaikina-Ruzmaikin conclude that the asymptotic form of the solution to (\ref{eq:beq}) is
$b(t) \approx \exp((t-t_0)^2 /12\GK)$ which  is obviously not approaching
the corresponding Friedmann solution as $t$ tends to infinity.

\label{sec:earlyTA}
We now move on to  perform a local dynamical systems analysis  in order to find the general
behaviour of the solutions to  Eq. (\ref{eq:beq}) near the initial singularity. This analysis is based on the use of the method
of asymptotic splittings
expounded in Ref. \cite{CB}. As a first
step, setting $b=x$, $\dot{b}=y$ and $\ddot{b}=z$, Eq. (\ref{eq:beq}) can be written as a dynamical system of the form
$\mathbf{\dot{x}}=\mathbf{f}(\mathbf{x})$:
\begin{equation}\label{eq:ds}
\dot{x} = y,\,\,\dot{y} = z,\,\,\dot{z} = \frac{y}{2\GK} - \frac{b_{1}^2}{2\GK yx^2} - \frac{yz}{x} + \frac{z^2}{2y} + \frac{3y^3}{2x^2}.
\end{equation}
If $\mathbf{a}=(\GA, \GB, \GG)$, and $\mathbf{p}=(p, q, r)$ we denote by $\mathbf{x}(\GT)$ the solution
\be
\mathbf{x}(\GT)=\mathbf{a}\GT^{\mathbf{p}}=(\GA \GT^{p}, \GB \GT^{q}, \GG \GT^{r})
\label{eq:domisol}
\ee
and by direct substitution to our system (\ref{eq:ds}) we look for the possible scale invariant
solutions\footnote{A vector field $\mathbf{f}$ is called scale invariant if
$\mathbf{f}(\mathbf{a}\GT^{\mathbf{p}})=\GT^{\mathbf{p-1}} \mathbf{f}(\mathbf{a})$}.
From all possible combinations, the most interesting is the one with \emph{dominant part} given by
\be
\mathbf{f}^{(0)}=\left(y,z,\frac{z^2}{2y}-\frac{zy}{x}+\frac{3y^3}{2x^2}  \right)
\label{eq:f0}
\ee
and \emph{subdominant part}
\be
\mathbf{f}^{\,\textrm{sub}}=\left(0,0,-\frac{b_{1}^{2}}{2\GK yx^2}+\frac{y}{2\GK}  \right),
\ee
where $\mathbf{f}=\mathbf{f}^{(0)} + \mathbf{f}^{\,\textrm{sub}}$. The dominant balance (of order 3) turns out to be
\be
(\mathbf{a},\mathbf{p}) = \left( \left( \GA,\frac{\GA}{2},-\frac{\GA}{4}\right),\: \left(\frac{1}{2},-\frac{1}{2},-\frac{3}{2} \right) \right)
\label{eq:domibal}
\ee
where $\GA$ is an arbitrary constant.

The Kowalevskaya exponents for this decomposition,
eigenvalues of the matrix
$\mathcal{K}=D\mathbf{f}(\mathbf{a})-\textrm{diag}(\mathbf{p})$, are $\{-1,0,3/2\}$
with corresponding eigenvectors $\{ (4,-2,3),(4,2,-1),(1,2,2)\}$.
The arbitrariness coming from the coefficient $\GA$ in the dominant balance reflects the fact that one of the
dominant exponents is zero with multiplicity one. According to the method of asymptotic splittings\cite{CB}, we proceed to construct
series expansions which are local solutions around movable singularities. In our particular problem the expansion around the singularity
turns out to be a Puiseux series of the form
\begin{equation} \label{eq:series}
x(t) = \sum_{i=0}^{\infty} c_{1i} (t-t_{0})^{\frac{i}{2}+\frac{1}{2}},\,
y(t) = \sum_{i=0}^{\infty} c_{2i}  (t-t_{0})^{\frac{i}{2}-\frac{1}{2}},\,
z(t) = \sum_{i=0}^{\infty} c_{3i} (t-t_{0})^{\frac{i}{2}-\frac{3}{2}} ,
\end{equation}
where $t_0$ is arbitrary and $c_{10}=\GA ,c_{20}=\GA /2,c_{30}=-\GA /4$. For these series expansions to be valid the compatibility condition
\be
(1,2,2) \cdot \left( \begin{array}{l}
                       -2c_{13}+c_{23}  \\
                       -c_{23}+c_{33}  \\
                       -\frac{1}{2}c_{13}+\frac{5}{4}c_{23}-c_{33}
                     \end{array}
              \right) = 0
\label{eq:cc}
\ee
must be satisfied. Substitution of Eq. (\ref{eq:series}) into Eq. (\ref{eq:ds}) leads to recursion relations that
determine the unknowns $c_{1i}, c_{2i}, c_{3i}$. After verifying that Eq. (\ref{eq:cc}) is indeed true, we write the final series
expansion corresponding to the balance (\ref{eq:domibal}). It is:
\be
x(t) = \GA \:\: (t-t_{0})^{\frac{1}{2}} + c_{13} \:\: (t-t_{0})^{2} + \displaystyle \frac{\GA^4-4b_{1}^2}{24\GK\GA^3} \:\:
                    (t-t_{0})^{\frac{5}{2}} + \cdots .
\label{eq:sol}
\ee
The series expansions for $y(t)$ and $z(t)$ are given by  the first and second time derivatives of the above expressions.

Our series (\ref{eq:sol})
has three arbitrary constants and is therefore  a local expansion of the \emph{general} solution around the movable singularity
$t_0$. Also since the leading order coefficients can be taken to be real, by a theorem of Goriely-Hyde \cite{GH}, we conclude that
 there is an open set of real initial conditions
for which the general solution blows up at the  (finite time) initial singularity at $t_0$. Finally, we observe that
near the initial singularity, the flat, radiation solutions of the higher order gravity theory considered here are Friedmann-like
regardless of the sign of the $R^2$ coefficient, while away from the singularity they strongly diverge from such forms.

\vspace{0.5cm}

\noindent This work was co-funded by 75\% from the EU and 25\% from the Greek Government, under the framework of the
``EPEAEK: Education and initial vocational training program - Pythagoras''.

\end{document}